\documentclass{article}

\usepackage{spconf,amsfonts,amsmath,graphicx,algorithmic,comment}

\begin{document}

\title{The Cascading Haar Wavelet algorithm for computing the Walsh-Hadamard Transform}
\name{Andrew Thompson}
\address{Mathematical Insititute\\University of Oxford\\United Kingdom}

\newtheorem{thm}{Theorem}
\newtheorem{lem}{Lemma}

\maketitle

\begin{abstract}
We propose a novel algorithm for computing the Walsh Hadamard Transform (WHT) which consists entirely of Haar wavelet transforms. We prove that the algorithm, which we call the Cascading Haar Wavelet (CHW) algorithm, shares precisely the same serial complexity as the popular divide-and-conquer algorithm for the WHT. We also propose a natural way of parallelizing the algorithm which has a number of attractive features.
\end{abstract}


\section{Introduction}

The Walsh-Hadamard Transform (WHT) is a staple of the digital signal processing world, and is used extensively in communication systems, image processing, and in general as a proxy for the Fast Fourier Transform (FFT)~\cite{walsh_paley}. Like the FFT, it is well known that the WHT of a signal of length $n$, where $n$ is a power of $2$, can be computed with $\mathcal{O}(n\log n)$ complexity. There exist well established algorithms for computing the WHT based on divide-and-conquer principles, which exploit the recursive properties of the transform, namely that any WHT of size $2^m$ can be broken down into two WHTs of size $2^{m-1}$. Various orderings of WHT coefficients are possible, most notably natural, dyadic and sequency orderings, and classical WHT algorithms essentially differ depending upon the desired ordering. See~\cite{walsh_paley} for background on the various WHT orderings and their corresponding algorithms. These fundamental algorithms have been known for many years, and more recent work has focused on practical considerations, such as how to incorporate these algorithms into parallel architectures and FPGAs; see for example~\cite{performance,FPGA}.

It has been previously noted that there exist interesting relationships between the WHT with dyadic ordering and the oldest and simplest discrete wavelet transform, the Haar wavelet transform. It was observed in~\cite{walsh_relations} that computing the WHT of a signal has a striking interpretation in terms of Haar wavelet coefficients: it is equivalent to applying WHTs of different sizes independently to the coefficients within each scale of the Haar wavelet transform. Based on this observation, the authors propose a \emph{Haar-Walsh} transform, which transforms Haar wavelet coefficients into WHT coefficients, thereby giving an alternative approach to computing the WHT: via a detour into the Haar wavelet domain. This approach was shown to match the complexity of the standard algorithms, and it has the additional appeal of computing the Haar wavelet transform for free in the process. Nonetheless, it appears that the approach never became a popular alternative to the standard WHT algorithms. We also note that, while a single Haar wavelet transform is computed at the start, the algorithm proceeds using the standard divide-and-conquer approach within each scale of the Haar wavelet transform thereafter.

In this paper, we propose a novel algorithm for computing the WHT with coefficients (in dyadic order) which is inspired by some of the connections between WHTs and Haar wavelet transforms, but which is fundamentally different from all preceding algorithms. Its marked difference is apparent from the fact that it consists entirely of Haar wavelet transforms. We show that the algorithm, which we call the Cascading Haar Wavelet (CHW) algorithm, matches the serial complexity of the standard algorithms for either the natural or dyadic orderings, requiring precisely $n\log_2 n$ addition operations for its computation. Furthermore, we propose a natural way of parallelizing the algorithm in such a way that each of the nodes in the parallel architecture performs a single fixed task, namely a Haar wavelet transform of a given size.

\section{Description of the algorithm}

Given $m\geq 0$, the $2^m\times 2^m$ Hadamard matrix with columns in dyadic (Paley) order~\cite{walsh_relations,walsh_paley}, $H_m$, is defined by the recursion
\begin{equation}\label{walsh_dyadic}
H_0:=1;\;\;\;\;H_{m+1}=\frac{1}{\sqrt{2}}\begin{bmatrix}\begin{array}{c}H_m\otimes\left(\begin{array}{ll}1&1\end{array}\right)\\ H_m\otimes\left(\begin{array}{ll}1&-1\end{array}\right)\end{array}\end{bmatrix}\;\mbox{for}\;m\geq 0,
\end{equation}
where $\otimes$ denotes the Kronecker product. Given $m\geq 0$, the $2^m\times 2^m$ Haar matrix~\cite{walsh_paley}, $\Psi_m$, may be defined by the recursion
\begin{equation}\label{haar}
\Psi_0:=1;\;\;\;\;\Psi_{m+1}=\frac{1}{\sqrt{2}}\begin{bmatrix}\begin{array}{c}\Psi_m\otimes\left(\begin{array}{ll}1&1\end{array}\right)\\I_m\otimes\left(\begin{array}{ll}1&-1\end{array}\right)\end{array}\end{bmatrix}\;\mbox{for}\;m\geq 0,
\end{equation}
where we write $I_m$ for the $2^m\times 2^m$ identity matrix.

The CHW algorithm is based on a particular decomposition of a Hadamard matrix in terms of Haar wavelet transform matrices. We use the notation $\displaystyle\prod_{r=1}^p M_r=M_p\cdots M_2 M_1$ for a $p$-fold matrix product.

\begin{thm}\label{decomp}
\begin{equation}\label{decomp_eqn}
H_m=\left\{\prod_{r=1}^{m-1}I_{r-1}\otimes\begin{bmatrix}I_{m-r}&0\\0&\Psi_{m-r}\end{bmatrix}\right\}\Psi_m,\;\mbox{for}\;m\geq 1.
\end{equation}
\end{thm}

\begin{figure}
\centering
\includegraphics[width=0.5\textwidth]{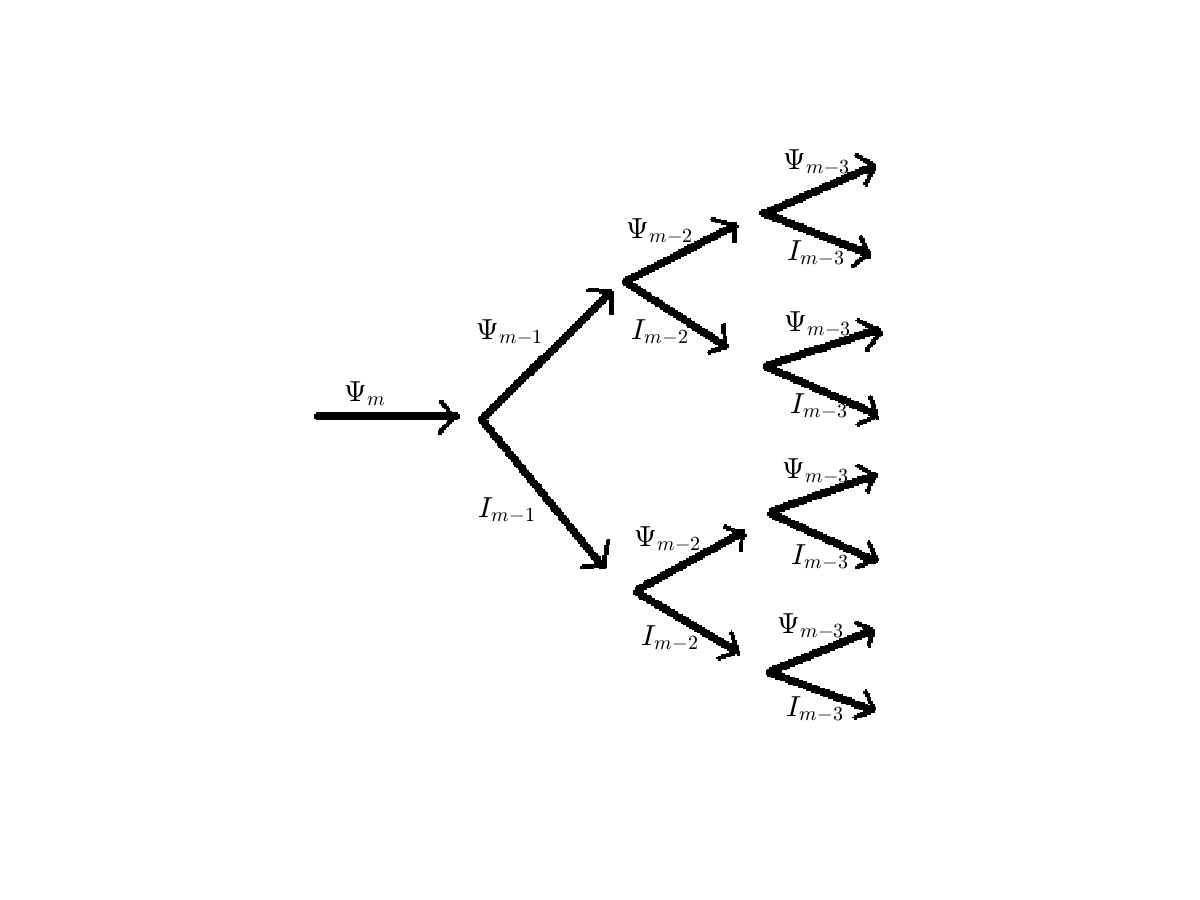}
\caption{An illustration of the Cascading Haar Wavelet algorithm.}
\label{flow_diagram}
\end{figure}

A proof of Theorem~\ref{decomp} is given in Section~\ref{derivation}. Expanding the product in (\ref{decomp_eqn}), we have
$$H_m=\left\{\begin{bmatrix}
I_1&&&&&&\\
&\Psi_1&&&&&\\
&&I_1&&&&\\
&&&\Psi_1&&&\\
&&&&\ddots&&\\
&&&&&I_1&\\
&&&&&&\Psi_1\end{bmatrix}\cdots\right.$$
$$\left.\cdots\begin{bmatrix}I_{m-2}&&&\\&\Psi_{m-2}&&\\&&I_{m-2}&\\&&&\Psi_{m-2}\end{bmatrix}\begin{bmatrix}I_{m-1}&\\&\Psi_{m-1}\end{bmatrix}\right\}\Psi_m,$$
which shows that the WHT can be computed by first computing the Haar wavelet transform, and then employing a divide-and-conquer approach also consisting of Haar wavelet transforms, as illustrated in Figure~\ref{flow_diagram}.

Figure~\ref{flow_diagram} is potentially misleading, in that the identity transforms do not actually need to be performed! We analyze the complexity of the CHW algorithm in Section~\ref{complexity}, where we show that the algorithm requires $n\log_2 n$ summations, where $n=2^m$ -- exactly the same as the standard WHT algorithms~\cite{walsh_paley}.

\section{Complexity analysis}\label{complexity}

We have proposed a method for computing the WHT which is built up entirely of Haar wavelet transforms. To analyze its complexity, we therefore need a complexity result for the Haar wavelet transform.

\begin{lem}[{\cite[Section 7.3.3]{orthogonal_transforms}}]\label{haar_lemma}
The Haar wavelet transform corresponding to multiplication by $\Psi_m$ can be computed in $2(2^m-1)$ operations.
\end{lem}

Equipped with this result, we can determine the complexity of the CHW algorithm.

\begin{thm}\label{CHW_complexity}
The CHW algorithm can be implemented in $m\cdot 2^m$ operations.
\end{thm}

\textbf{Proof:} From Figure~\ref{flow_diagram} we see that the CHW algorithm requires a single Haar wavelet transform of size $2^m$, and $2^{m-1-r}$ Haar wavelet transforms of size $2^r$, for $r=1,2,\ldots,m-1$. By Lemma~\ref{haar_lemma}, the total number of operations is therefore
\begin{eqnarray}
&&2(2^m-1)+\sum_{r=1}^{m-1}\left\{2^{m-1-r}\cdot 2(2^r-1)\right\}\nonumber\\
&=&2^{m+1}-2+\sum_{r=1}^{m-1}2^m-\sum_{r=1}^{m-1}2^{m-r}\nonumber\\
&=&2^{m+1}-2+2^m(m-1)-2(2^{m-1}-1),\nonumber
\end{eqnarray}
which simplifies to $m\cdot 2^m$.\hfill$\Box$

We have shown that the CHW algorithm has precisely the same serial complexity as the popular divide-and-conquer algorithms for the WHT. In the next section, we propose a natural way of parallelizing the CHW which has a number of attractive features.

\section{A proposal for a parallel implementation}

Given the WHT's importance in signal processing, it is not surprising that there already exists a body of work addressing the question of how to efficiently parallelize it; see~\cite{performance} for an example. In this section, we make the observation that there is a very natural way to parallelize the CHW algorithm, which possesses a number of attractive features.

\begin{figure}
\centering
\includegraphics[width=0.5\textwidth]{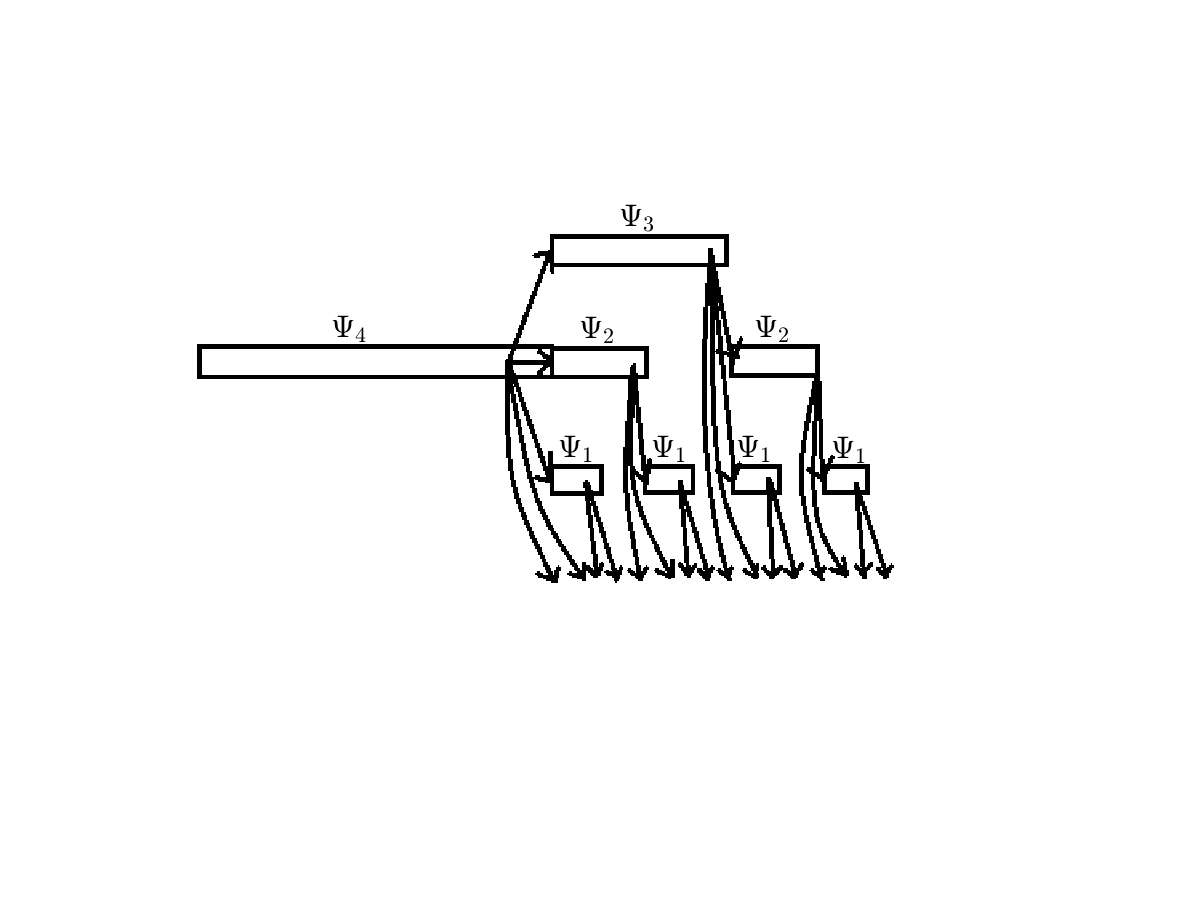}
\caption{An illustration of the proposed parallel implementation of the CHW algorithm.}
\label{parallel_flow}
\end{figure}

In the CHW algorithm, a signal of length $2^m$ is cascaded through a succession of Haar wavelet transforms. It is possible therefore to consider a parallel architecture in which each of $m-1$ nodes is devoted to the task of performing Haar wavelet transforms of a certain size. A scheduling chart illustrating this procedure for $m=4$ is shown in Figure~\ref{parallel_flow}. In this case, we have three nodes, each devoted to the task of performing the Haar wavelet transforms $\Psi_1$, $\Psi_2$ and $\Psi_3$. A full Haar wavelet transform $\Psi_4$ must first be performed (by one of the three nodes, or by an extra one), and thereafter each node is occupied for approximately half of the total running time. The output is the WHT coefficients in dyadic order. Note the attractive properties of this scheme: each node need only be programmed to perform a single task, and communication of the output from any given node follows fixed and straightforward rules. 

\section{Proof of Theorem 1}\label{derivation}

We proceed by induction. The result holds trivially for $m=1$. Assume (\ref{decomp_eqn}) holds for $m-1$. Then
\begin{eqnarray}
&&\sqrt{2}\left\{\prod_{r=1}^{m-1}I_{r-1}\otimes\begin{bmatrix}I_{m-r}&0\\0&\Psi_{m-r}\end{bmatrix}\right\}\Psi_m\nonumber\\
&=&\left\{\prod_{r=1}^{m-1}I_{r-1}\otimes\begin{bmatrix}I_{m-r}&0\\0&\Psi_{m-r}\end{bmatrix}\right\}\begin{bmatrix}\begin{array}{c}\Psi_m\otimes\left(\begin{array}{ll}1&1\end{array}\right)\\I_m\otimes\left(\begin{array}{ll}1&-1\end{array}\right)\end{array}\end{bmatrix}\nonumber
\end{eqnarray}
\begin{eqnarray}
&=&\left\{\prod_{r=2}^{m-1}I_{r-1}\otimes\begin{bmatrix}I_{m-r}&0\\0&\Psi_{m-r}\end{bmatrix}\right\}\nonumber\\
&&\;\;\;\;\cdot\begin{bmatrix}I_{m-1}&0\\0&\Psi_{m-1}\end{bmatrix}\begin{bmatrix}\begin{array}{c}\Psi_{m-1}\otimes\left(\begin{array}{ll}1&1\end{array}\right)\\I_{m-1}\otimes\left(\begin{array}{ll}1&-1\end{array}\right)\end{array}\end{bmatrix}\nonumber\\
&=&\left\{\prod_{r=2}^{m-1}I_{r-1}\otimes\begin{bmatrix}I_{m-r}&0\\0&\Psi_{m-r}\end{bmatrix}\right\}\begin{bmatrix}\begin{array}{c}\Psi_{m-1}\otimes\left(\begin{array}{ll}1&1\end{array}\right)\\\Psi_{m-1}\otimes\left(\begin{array}{ll}1&-1\end{array}\right)\end{array}\end{bmatrix}\nonumber\\
&=&I_1\otimes\left\{\prod_{r=2}^{m-1}I_{r-2}\otimes\begin{bmatrix}I_{m-r}&0\\0&\Psi_{m-r}\end{bmatrix}\right\}\nonumber\\
&&\;\;\;\;\cdot\begin{bmatrix}\begin{array}{c}\Psi_{m-1}\otimes\left(\begin{array}{ll}1&1\end{array}\right)\\\Psi_{m-1}\otimes\left(\begin{array}{ll}1&-1\end{array}\right)\end{array}\end{bmatrix}\nonumber\\
&=&\begin{bmatrix}\begin{array}{c}\displaystyle\prod_{r=1}^{m-2}I_{r-1}\otimes\begin{bmatrix}I_{m-1-r}&0\\0&\Psi_{m-1-r}\end{bmatrix}\Psi_{m-1}\otimes\left(\begin{array}{ll}1&1\end{array}\right)\\ \displaystyle\prod_{r=1}^{m-2}I_{r-1}\otimes\begin{bmatrix}I_{m-1-r}&0\\0&\Psi_{m-1-r}\end{bmatrix}\Psi_{m-1}\otimes\left(\begin{array}{ll}1&-1\end{array}\right)\end{array}\end{bmatrix},\nonumber
\end{eqnarray}
which, by the inductive hypothesis, is equal to
\begin{eqnarray}
&=&\begin{bmatrix}H_{m-1}\Psi^T_{m-1}\Psi_{m-1}\otimes\left(\begin{array}{ll}1&1\end{array}\right)\\H_{m-1}\Psi^T_{m-1}\Psi_{m-1}\otimes\left(\begin{array}{ll}1&-1\end{array}\right)\end{bmatrix}\nonumber\\
&=&\begin{bmatrix}H_m\otimes\left(\begin{array}{ll}1&1\end{array}\right)\\ H_m\otimes\left(\begin{array}{ll}1&-1\end{array}\right)\end{bmatrix},\nonumber\\
&=&\sqrt{2}H_m.\nonumber
\end{eqnarray}

\section{Relation to prior work}

 The author is aware of two papers especially in which the relationship between the WHT and the Haar wavelet transform has been explored. In~\cite{walsh_relations}, the authors consider a \emph{Haar-Walsh} transform which transforms Haar wavelet coefficients into WHT coefficients (in dyadic order), which they observe to be equivalent to multiplication by the matrix
\begin{equation}\label{decomp2}
H_m\Psi_m^T=\begin{bmatrix}
1&&&&\\
&H_0&&&\\
&&H_1&&\\
&&&\ddots&\\
&&&&H_{m-1}\end{bmatrix}.\end{equation}
The Haar-Walsh transform can therefore be computed by taking separate WHTs of the Haar wavelet coefficients at each scale. See also~\cite{multilevel} by the current author in which the implications of this decomposition are explored for multilevel compressive sensing.

The possibility of recursively decomposing Hadamard matrices using (\ref{decomp2}) appears to be spotted in the concluding remarks of~\cite{fino}, and indeed it is possible to derive the CHW algorithm from repeated application of (\ref{decomp2}). Closely related though the ideas in~\cite{fino} are, to the author's best knowledge, there is no mention in the literature of a WHT consisting entirely of Haar wavelet transforms, nor any statement of the decomposition result given in Theorem~\ref{decomp}.

\section{Concluding remarks}

We have proposed the novel Cascading Haar Wavelet (CHW) algorithm for computing the WHT. We have also proposed a parallelization scheme, and it remains to comprehensively understand the practical implementation advantages that the CHW might have over other approaches to parallelization of the WHT.

\end{document}